\documentclass[english,aps,prl,twocolumn,floatfix,showpacs,superscriptaddress]{revtex4}
\usepackage[T1]{fontenc}
\usepackage[latin9]{inputenc}
\usepackage{color}
\usepackage{babel}
\usepackage{amsmath}
\usepackage{graphicx}
\usepackage{esint}
\usepackage[unicode=true,pdfusetitle,
 bookmarks=true,bookmarksnumbered=false,bookmarksopen=false,
 breaklinks=false,pdfborder={0 0 0},backref=false,colorlinks=true]
 {hyperref}

\makeatletter

\newcommand{\lyxdot}{.}

\@ifundefined{textcolor}{}
{%
 \definecolor{BLACK}{gray}{0}
 \definecolor{WHITE}{gray}{1}
 \definecolor{RED}{rgb}{1,0,0}
 \definecolor{GREEN}{rgb}{0,1,0}
 \definecolor{BLUE}{rgb}{0,0,1}
 \definecolor{CYAN}{cmyk}{1,0,0,0}
 \definecolor{MAGENTA}{cmyk}{0,1,0,0}
 \definecolor{YELLOW}{cmyk}{0,0,1,0}
}

\makeatother

\begin{document}

\title{Mott physics and spin fluctuations: a unified framework}

\author{Thomas Ayral}

\email{thomas.ayral@polytechnique.edu}

\selectlanguage{english}%

\affiliation{Centre de Physique Théorique, Ecole Polytechnique, CNRS-UMR7644,
91128 Palaiseau, France}

\affiliation{Institut de Physique Théorique (IPhT), CEA, CNRS, URA 2306, 91191
Gif-sur-Yvette, France}

\author{Olivier Parcollet}

\affiliation{Institut de Physique Théorique (IPhT), CEA, CNRS, URA 2306, 91191
Gif-sur-Yvette, France}
\begin{abstract}
We present a formalism for strongly correlated electrons systems which
consists in a local approximation of the dynamical three-leg interaction
vertex. This vertex is self-consistently computed with a quantum impurity
model with dynamical interactions in the charge and spin channels,
similar to dynamical mean field theory (DMFT) approaches. The electronic
self-energy and the polarization are both frequency and momentum dependent.
The method interpolates between the spin-fluctuation or GW approximations
at weak coupling and the atomic limit at strong coupling. We apply
the formalism to the Hubbard model on a two-dimensional square lattice
and show that as interactions are increased towards the Mott insulating
state, the local vertex acquires a strong frequency dependence, driving
the system to a Mott transition, while at low enough temperatures
the momentum-dependence of the self-energy is enhanced due to large
spin fluctuations. Upon doping, we find a Fermi arc in the one-particle
spectral function, which is one signature of the pseudo-gap state.
\end{abstract}
\maketitle
Strongly-correlated electronic systems like high-temperature cuprate
superconductors are a major challenge in condensed-matter physics. 

One theoretical approach to cuprates emphasizes the effect of long-range
bosonic fluctuations on the electronic fluid, for example long-range
antiferromagnetic (AF) fluctuations due to a quantum critical point
\cite{Chubukov2002,Efetov2013,Wang2014,Metlitski2010,Onufrieva2009,Onufrieva2012}.
These bosonic fluctuations are also central to approaches such as
the two-particle self-consistent approximation (TPSC \cite{Vilk1994,Dare1996,Vilk1996,Vilk1997,Tremblay2012}),
the GW approximation \cite{Hedin1965} and the fluctuation-exchange
approximation (FLEX \cite{Bickers1989}). 

Another approach focusses, following Anderson \cite{Anderson1987},
on describing the Mott transition and the doped Mott insulator. In
recent years, dynamical mean field theory (DMFT) \cite{Georges1996}
and its cluster extensions like CDMFT \cite{Lichtenstein2000,Kotliar2001}
or DCA \cite{Hettler1998,Hettler1999,Maier2005a} have allowed for
tremendous theoretical progress on the Mott transition both for models
and realistic computations of strongly correlated materials \cite{Kotliar2006}.
In particular, numerous works have been devoted to the one-band Hubbard
model, mapping out its phase diagram, studying the $d$-wave superconducting
order and the pseudogap \cite{Kyung2009,Sordi2012,Civelli2008,Ferrero2010,Gull2013,Macridin2004,Maier2004,Maier2005,Maier2006,Gull2010,Yang2011,Macridin2008,Macridin2006,Jarrell2001,Bergeron2011,Kyung2004,Kyung2006a,Okamoto2010,Sordi2010,Sordi2012a,Civelli2005,Ferrero2008,Ferrero2009,Gull2009}.
Cluster DMFT is one of the few methods designed for the strong-interaction
regime to have a simple control parameter, namely the size $N_{c}$
of the cluster or the momentum resolution of the electronic self-energy.
It interpolates between the DMFT solution ($N_{c}=1$) and the exact
solution of the Hubbard model ($N_{c}=\infty$). Despite its success,
this method nonetheless suffers from severe limitations: \emph{i)
}it does not include the effect of long-range bosonic modes of wavelengths
larger than the cluster size; \emph{ii)} the negative sign problem
of continuous-time quantum Monte Carlo has so far precluded the convergence
of the cluster solutions with respect to $N_{c}$ in the most important
regimes like the pseudogap; \emph{iii) }the $\mathbf{k}$-resolution
of the self-energy is still quite coarse in DCA (typically 8 or 16
patches in the Brillouin zone, see e.g. \cite{Gull2009,Gull2010,Vidhyadhiraja2009,Macridin2008}),
or relies on uncontrolled a posteriori ``periodization'' techniques
in CDMFT \cite{Kotliar2001}.

Several directions beyond cluster DMFT methods are currently under
investigation to address these issues, such as GW + DMFT \cite{Sun2002,Sun2008,Biermann2003,Ayral2012,Ayral2013,Hansmann2013,Huang2014},
the $\mathrm{D\Gamma A}$ method \cite{Toschi2007,Katanin2009,Schafer2014,Valli2014},
the dual fermion \cite{Rubtsov2008} and dual boson methods \cite{Rubtsov2011,Hafermann2014},
or combinations of DMFT with functional renormalization group methods
\cite{Taranto2014}.

In this letter, we discuss a simple formalism that unifies the two
points of view mentioned above. It is designed to encompass both Mott
physics \emph{à la }DMFT\emph{ }and the effect of medium and long-range
bosonic modes. It interpolates between the atomic limit in the strong-interaction
regime and the ``fluctuation-exchange'' limit in the weak-interaction
regime. It consists in decoupling the electron-electron interaction
term by Hubbard-Stratonovich bosonic fields and making a local self-consistent
approximation of the lattice's electron-boson one-particle irreducible
vertex, using a quantum impurity model similar to the one used in
DMFT. It can be formally derived from a functional of the vertex given
by three-particle irreducible diagrams \cite{Dominicis1964,Dominicis1964a}.
In the following, we will therefore denote this method as a triply-irreducible
local expansion, or TRILEX. Already at the single-site level, it produces,
in some parameter regimes, a momentum-dependent self-energy and polarization,
at a small computational cost, similar to solving Extended DMFT (EDMFT)
\cite{Sengupta1995,Kajueter1996,Si1996}. In the following, we first
introduce the method; we then present the solution of the single-site
version of TRILEX for the two-dimensional Hubbard model.

\begin{figure}
\begin{centering}
\includegraphics[scale=0.65]{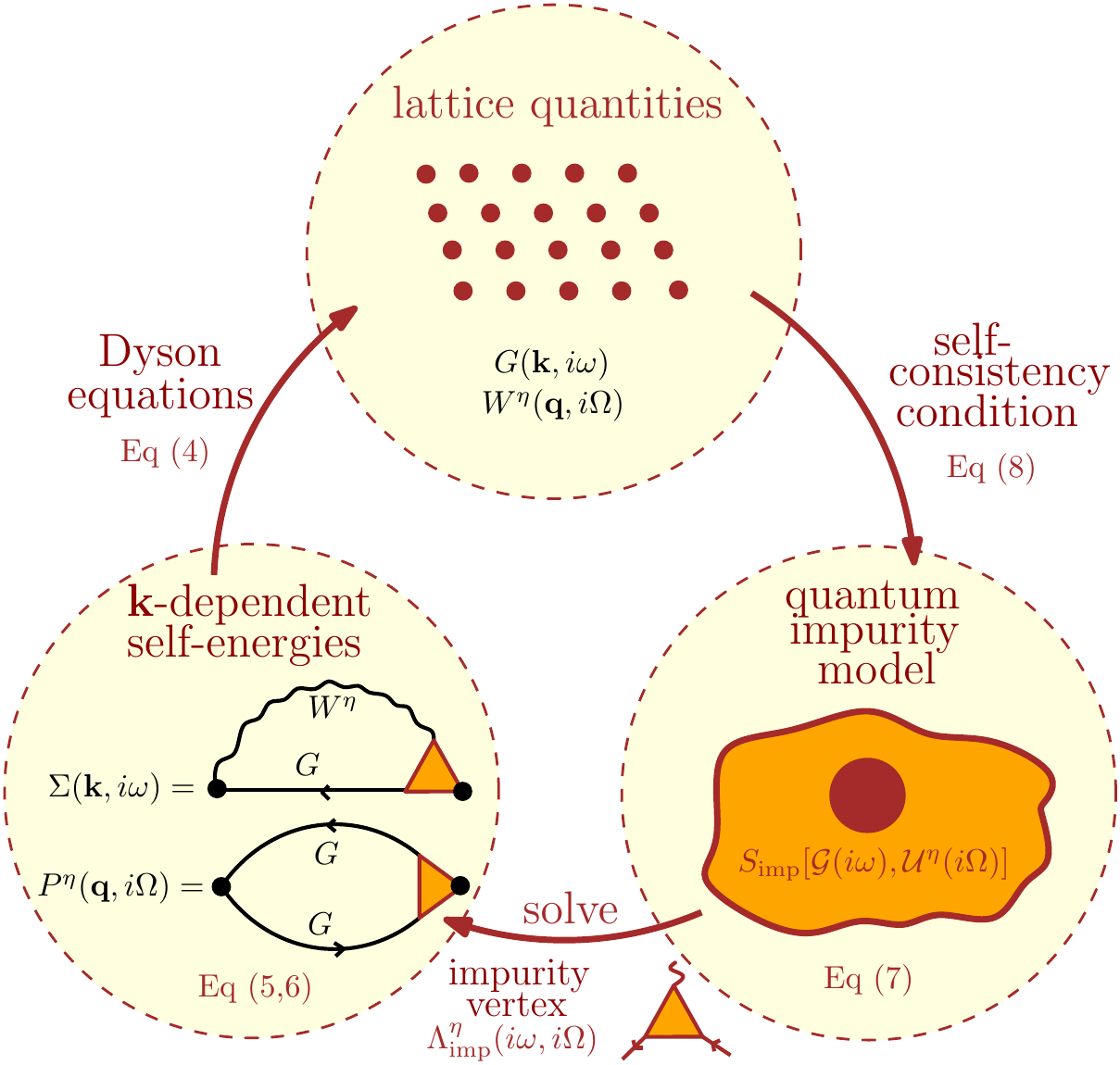}
\par\end{centering}

\protect\caption{(color online) Description of the TRILEX method \label{fig:scheme_sketch}}
\end{figure}

We focus on the Hubbard model defined by the following Hamiltonian:
\begin{equation}
H=\sum_{ij}t_{ij}c_{i}^{\dagger}c_{j}+U\sum_{i}n_{i\uparrow}n_{i\downarrow}\label{eq:Hubbard_model}
\end{equation}
The indices $i,j$ denote lattice sites, $\sigma=\uparrow,\downarrow$,
$c_{i\sigma}^{\dagger}$ and $c_{i\sigma}$ are fermionic creation
and annihilation operators, and $n_{i\sigma}\equiv c_{i\sigma}^{\dagger}c_{i\sigma}$.
$t_{ij}$ is the tight-binding hopping matrix $(t_{ij}=t(t')$ for
(next-)nearest-neighbors), while $U$ is the on-site Coulomb interaction.
We rewrite the operators of the interaction term as: 
\begin{equation}
Un_{i\uparrow}n_{i\downarrow}=\frac{1}{2}\sum_{I}U^{I}n_{i}^{I}n_{i}^{I}\label{eq:channel_decomposition}
\end{equation}
where $U^{I}$ is the bare interaction in channel $I$, and $n_{i}^{I}\equiv\sum_{\sigma\sigma'}c_{i\sigma}^{*}\sigma_{\sigma\sigma'}^{I}c_{i\sigma'}$
where $\sigma^{0}=\mathbf{1}$ and $\sigma^{x,y,z}$ are the Pauli
matrices. In this paper, we consider two decouplings: (a) in the charge
and vector spin channel ($I=0,x,y,z$) (``$xyz$-decoupling''),
$U^{x}=U^{y}=U^{z}\equiv U^{\mathrm{sp}}$ and $U^{0}\equiv U^{\mathrm{ch}}$,
$U^{\mathrm{sp}}$ and $U^{\mathrm{ch}}$ satisfy: $U=U^{\mathrm{ch}}-3U^{\mathrm{sp}}$;
(b) in the charge and longitudinal spin channel only ($I=0,z$) (``$z$-decoupling''),
$U=U^{\mathrm{ch}}-U^{\mathrm{sp}}$. In both cases, we have two channels,
denoted as $\eta=\mathrm{ch},\mathrm{sp}$. In this paper, we fix
the ratio to $U^{\mathrm{ch}}=U/2$ and $U^{\mathrm{sp}}=-U/6$ ($xyz$
decoupling) and $U^{\mathrm{ch}}=U/2$ and $U^{\mathrm{sp}}=-U/2$
($z$ decoupling). We now decouple (\ref{eq:channel_decomposition})
using real bosonic Hubbard-Stratonovich fields $\phi_{i}^{I}(\tau)$
in each channel and at each lattice site, so that the action becomes:

\begin{eqnarray}
S_{\mathrm{latt}} & = & \int_{0}^{\beta}d\tau\sum_{ij}c_{i\sigma\tau}^{*}\left\{ \partial_{\tau}+t_{ij}\right\} c_{j\sigma\tau}\nonumber \\
 &  & +\sum_{i,I}\left[\frac{1}{2}(U^{I})^{-1}\phi_{i\tau}^{I}\phi_{i\tau}^{I}+\lambda^{I}\phi_{i\tau}^{I}n_{i\tau}^{I}\right]\label{eq:electron-boson_action}
\end{eqnarray}
$c_{i\sigma\tau}^{*}$ and $c_{i\sigma\tau}$ are conjugate $\beta$-antiperiodic
Grassmann fields, and $\lambda^{I}=1$. We are now dealing with an
interacting lattice problem with a local electron-boson coupling.
The lattice Green's functions $G(\mathbf{k},i\omega)$ and $W^{\eta}(\mathbf{q},i\Omega)$
(the Fourier transforms of $-\langle c_{i\sigma\tau}c_{j\sigma0}^{*}\rangle$
and $-\langle\phi_{i\sigma\tau}^{\eta}\phi_{j\sigma0}^{\eta}\rangle$,
respectively) are given by Dyson equations:

\begin{subequations}

\begin{eqnarray}
G(\mathbf{k},i\omega) & = & \left[i\omega+\mu-\epsilon(\mathbf{k})-\Sigma(\mathbf{k},i\omega)\right]^{-1}\label{eq:Dyson_G}\\
W^{\eta}(\mathbf{q},i\Omega) & = & U^{\eta}\left[1-U^{\eta}P^{\eta}(\mathbf{q},i\Omega)\right]^{-1}\label{eq:Dyson_W}
\end{eqnarray}
\end{subequations}$\mathbf{k}$ and $\mathbf{q}$ are momentum variables,
$i\omega$($i\Omega$) stands for a fermionic (bosonic) Matsubara
frequency, $\epsilon(\mathbf{k})$ is the Fourier transform of $t_{ij}$,
and $\mu$ is the chemical potential. The fermionic and bosonic self-energies
$\Sigma$ and $P^{\eta}$ are given by the exact expressions (written
here for the paramagnetic normal phase) (see \emph{e.g }\cite{Aryasetiawan2008}):

\begin{subequations}

\begin{eqnarray}
\Sigma(\mathbf{k},i\omega) & = & -\sum_{\substack{\mathbf{q},i\Omega,\\
\eta=\mathrm{ch},\mathrm{sp}
}
}m_{\eta}\lambda^{\eta}G_{\substack{\mathbf{q+k},\\
i\omega+i\Omega
}
}W_{\mathbf{q},i\Omega}^{\eta}\Lambda_{\substack{\mathbf{k},\mathbf{q},\\
i\omega,i\Omega
}
}^{\eta}\label{eq:Sigma_exact}\\
P^{\eta}(\mathbf{q},i\Omega) & = & 2\sum_{\mathbf{k},i\omega}\lambda^{\eta}G_{\substack{\mathbf{q+k},\\
i\omega+i\Omega
}
}G_{\mathbf{k},i\omega}\Lambda_{\substack{\mathbf{k},\mathbf{q},\\
i\omega,i\Omega
}
}^{\eta}\label{eq:P_exact}
\end{eqnarray}
\end{subequations}Here, $m_{\mathrm{ch}}=1$, $m_{\mathrm{sp}}=3$
($xyz$ decoupling) or $m_{\mathrm{sp}}=1$ ($z$ decoupling). $\Lambda^{\eta}(\mathbf{q},\mathbf{k},i\omega,i\Omega)$
is the exact one-particle irreducible electron-boson coupling (or
Hedin) vertex, namely the effective interaction between electrons
and bosons renormalized by electronic interactions. 

The main point of this paper consists in approximating the vertex
$\Lambda^{\eta}(\mathbf{q},\mathbf{k},i\omega,i\Omega)$ by the local,
but two-frequency-dependent $\Lambda_{\mathrm{imp}}^{\eta}(i\omega,i\Omega)$
computed from a self-consistent quantum impurity problem:
\begin{equation}
\Lambda^{\eta}(\mathbf{q},\mathbf{k},i\omega,i\Omega)\approx\Lambda_{\mathrm{imp}}^{\eta}(i\omega,i\Omega)\label{eq:TRILEX_approximation}
\end{equation}
This strategy radically differs from DMFT, EDMFT and GW+DMFT which
approximate the self-energy $\Sigma$ (and $P$), not $\Lambda$.
It implies that our $\Sigma$ and $P$ (computed from (\ref{eq:Sigma_exact}-\ref{eq:P_exact}))
are, in some parameter regimes, strongly \emph{momentum-dependent}
while containing \emph{local} vertex corrections which will be essential
to capture Mott physics (see also \cite{Ayral2012}). Formally, DMFT
is a local approximation of the two-particle irreducible Luttinger-Ward
functional $\Phi_{\mathrm{LW}}[G]$ \cite{Georges1991,Georges1996}.
In contrast, our approximation can be defined as a local approximation
of the \emph{three-particle} irreducible functional $\mathcal{K}[G,W,\Lambda]$
introduced in \cite{Dominicis1964,Dominicis1964a} as a generalization
of $\Phi_{\mathrm{LW}}$ to higher degrees of irreducibility. We therefore
denote it as TRILEX, triply-irreducible local expansion. It makes
it exact in the limit of infinite dimensions. The formal derivation
of the method will be provided elsewhere \cite{Ayral}. 

\begin{figure}
\begin{centering}
\includegraphics[scale=0.46]{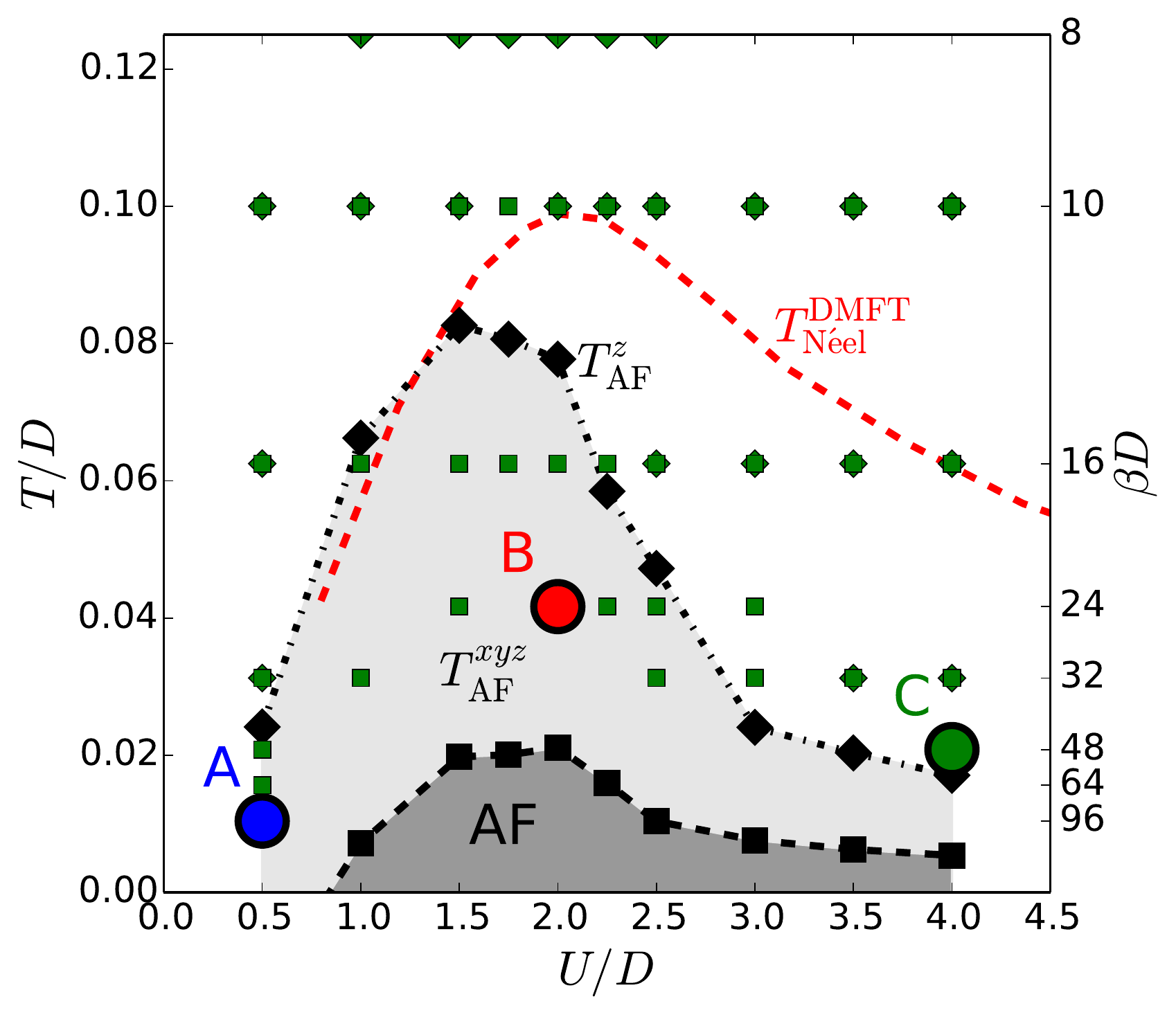}
\par\end{centering}

\protect\caption{\label{fig:TU_Phase-diagram}(color online) $(T,U$) Phase diagram
(half-filling, $t'=0$). The black diamonds (resp. squares) denote
$T_{\mathrm{AF}}^{z}$ (resp. $T_{\mathrm{AF}}^{xyz}$); the dashed
lines are guides to the eye. The green diamonds (resp. squares) denote
converged TRILEX solutions in the $z$ (resp. $x,y,z$) decoupling.
A, B, and C are defined as A: $\beta D=96$, $U/D=0.5$, B: $\beta D=24$,
$U/D=2$, C: $\beta D=48$, $U/D=4$. The red dotted line denotes
$T_{\mathrm{N\acute{e}el}}^{DMFT}$ for the square lattice (from \cite{Kunes2011}).}
\end{figure}

The action of the impurity model reads:
\begin{eqnarray}
S_{\mathrm{imp}} & = & -\iint_{0}^{\beta}d\tau d\tau'\sum_{\sigma}c_{\sigma\tau}^{*}\mathcal{G}(\tau-\tau')c_{\sigma\tau'}\nonumber \\
 &  & +\frac{1}{2}\sum_{I}\iint_{0}^{\beta}d\tau d\tau'n_{\tau}^{I}\mathcal{U}^{I}(\tau-\tau')n_{\tau'}^{I}\label{eq:local_effective_action}
\end{eqnarray}
This is an Anderson quantum impurity with retarded charge-charge ($I=0$)
and spin-spin ($I=x,y,z$ in the $xyz$-decoupling, $I=z$ in the
$z$-decoupling) interactions. The bosonic fields $\phi^{I}$ have
been integrated out to obtain a fermionic action with retarded interactions
amenable to numerical computations. We compute the fermionic three-point
correlation functions to reconstruct the electron-boson vertex $\Lambda_{\mathrm{imp}}$
(as shown in the Suppl. Mat., section \ref{sec:Computation-of-the-three-leg-vertex}).
Finally, $\mathcal{G}$ and $\mathcal{U}^{\eta}$ derive from the
self-consistency conditions as follows:

\begin{subequations}

\begin{eqnarray}
\mathcal{G}^{-1}(i\omega) & = & G_{\mathrm{loc}}^{-1}(i\omega)+\Sigma_{\mathrm{loc}}(i\omega)\label{eq:G_weiss}\\
\left[\mathcal{U}^{\eta}\right]^{-1}(i\omega) & = & \left[W_{\mathrm{loc}}^{\eta}\right]^{-1}(i\omega)+P_{\mathrm{loc}}^{\eta}(i\omega)\label{eq:U_weiss}
\end{eqnarray}
\end{subequations}where, for any $X$, $X_{\mathrm{loc}}(i\omega)\equiv\sum_{\mathbf{k}}X(\mathbf{k},i\omega)$.
At convergence, this ensures that $G_{\mathrm{loc}}=G_{\mathrm{imp}}$
and $W_{\mathrm{loc}}^{\eta}=W_{\mathrm{imp}}^{\eta}$. $W^{\eta}$
and the susceptibility $\chi^{\eta}$ are related by:

\begin{equation}
W^{\eta}(\mathbf{q},i\Omega)=U^{\eta}-U^{\eta}\chi^{\eta}(\mathbf{q},i\Omega)U^{\eta}\label{eq:W_VS_chi}
\end{equation}

The computational scheme is illustrated in Fig. \ref{fig:scheme_sketch}.
From the impurity electron-boson vertex $\Lambda_{\mathrm{imp}}$,
we compute $\Sigma(\mathbf{k},i\omega)$ and $P^{\eta}(\mathbf{q},i\Omega)$,
which are then used to compute $\mathcal{G}$ and $\mathcal{U}^{\eta}$
for (\ref{eq:local_effective_action}). We solve the quantum impurity
model exactly by a continuous-time quantum Monte-Carlo algorithm \cite{Rubtsov2005}
in the hybridization expansion \cite{Werner2006} with retarded density-density
\cite{Werner2007} and vector spin-spin interactions \cite{Otsuki2013}.
The computation of the three-point functions are implemented as described
in \cite{Hafermann2013}. We iterate until convergence is reached.
Our implementation is based on the TRIQS library \cite{Parcollet2014}. 

TRILEX provides a unified framework for spin-fluctuation approaches
and Mott physics. Indeed, (i) at small interaction strengths, the
local vertex reduces to the bare, frequency-independent vertex $\lambda^{\eta}$
so that $\Sigma$ is given by one-loop self-consistent diagrams, as
in spin fluctuation theory in its simplest form (spin channel only),
the GW approximation (charge channel only), or in FLEX limited to
particle-hole diagrams; similarly, $P^{\eta}$ becomes equal to the
``bubble'' diagram; (ii) it is exact in the atomic limit ($t=0$):
the effective local action turns into an atomic problem, $\Lambda$
into the atomic vertex $\Lambda_{\mathrm{at}}$ (Eq.\ref{eq:Lambda_at}),
and $\Sigma$ and $P$ become local, atomic self-energies.

\begin{figure}
\begin{centering}
\includegraphics[scale=0.36]{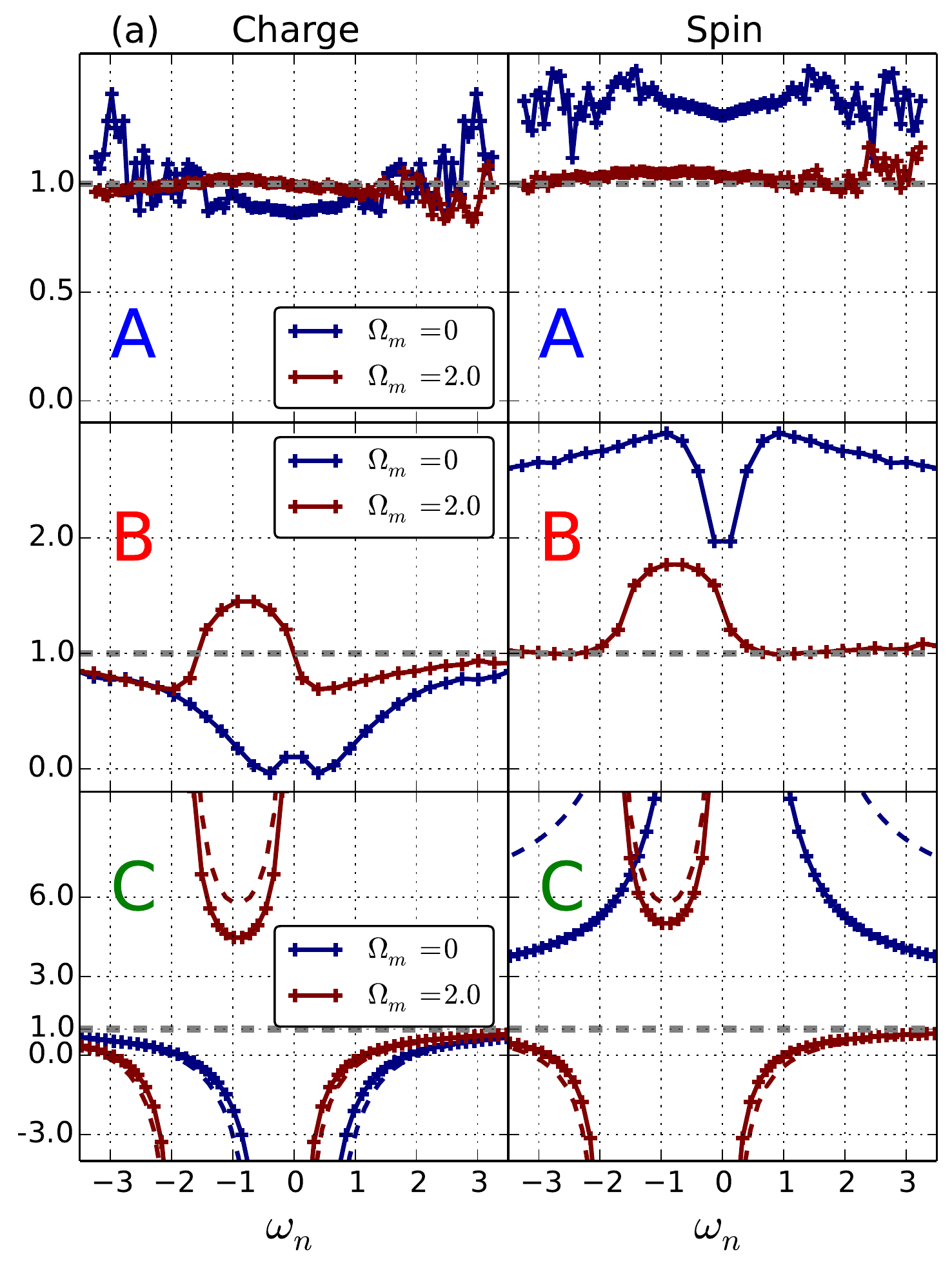}\includegraphics[scale=0.39]{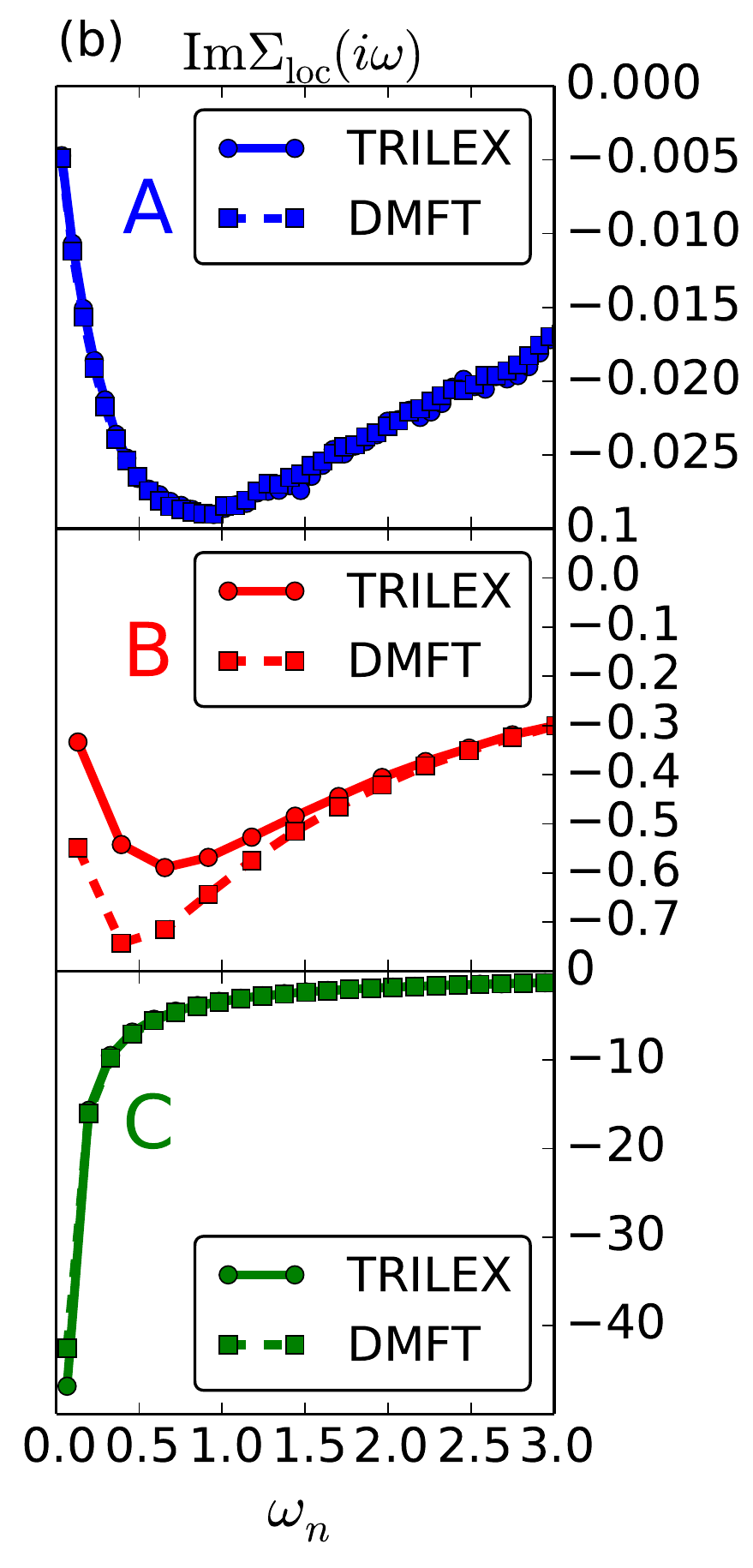}
\par\end{centering}

\protect\caption{\label{fig:local_vertex}(color online) Left: Evolution of the local
vertex $\mathrm{Re}\Lambda^{\eta}(i\omega_{n},i\Omega_{m})$ (half-filling,
$t'=0$). A, B and C are defined in Fig. \ref{fig:TU_Phase-diagram}.
The dashed lines denote the atomic vertex $\Lambda_{\mathrm{at}}^{\eta}$
(Eq. (\ref{eq:Lambda_at})). Right: $\mathrm{Im}\Sigma_{\mathrm{loc}}(i\omega_{n})$
for TRILEX and DMFT (paramagnetic phase).}
\end{figure}

\begin{figure}
\begin{centering}
\includegraphics[scale=0.33]{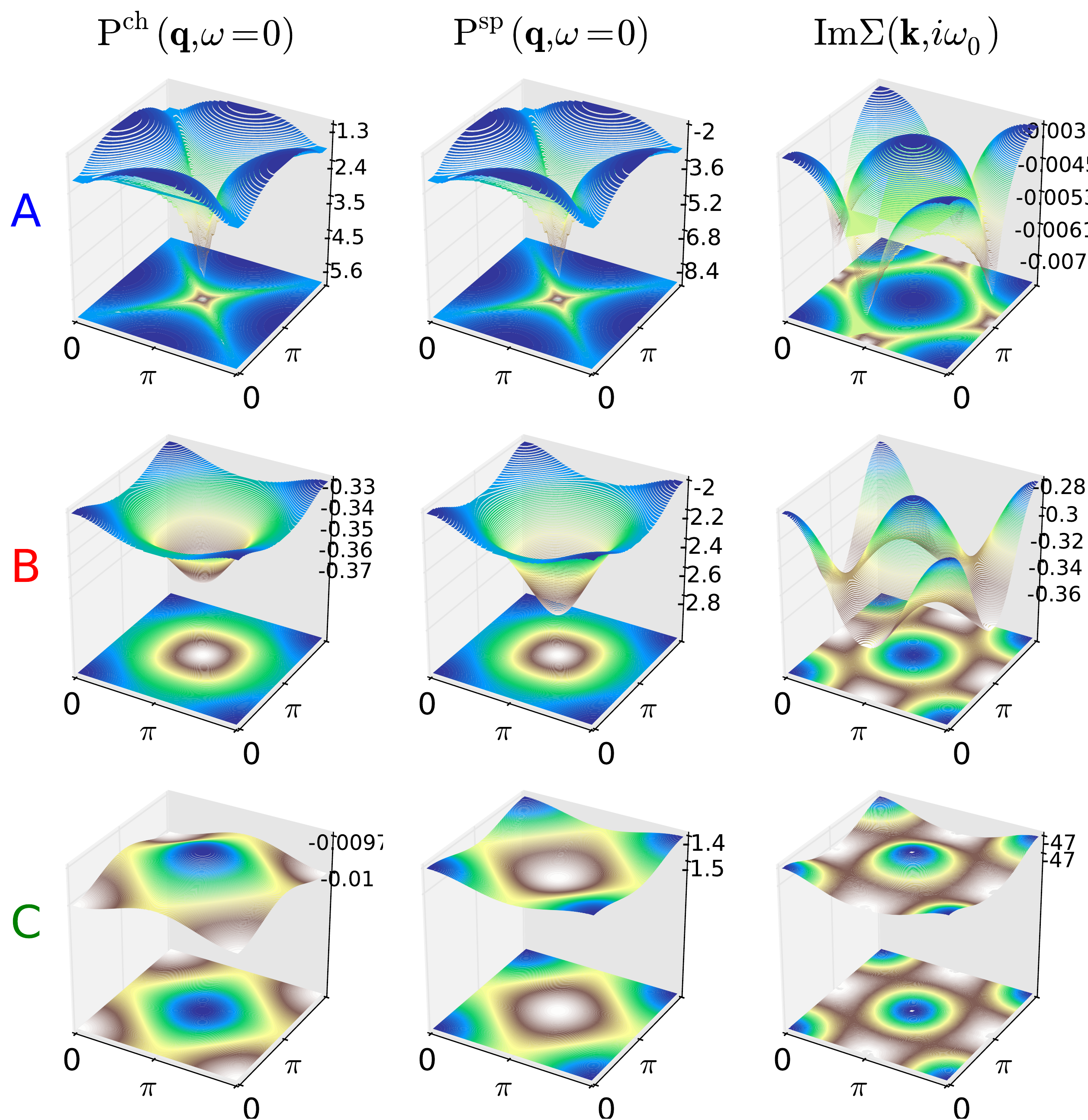}
\par\end{centering}

\protect\caption{(color online)\label{fig:Momentum-dependence-Self-energies} Momentum-dependence
of the self-energy and polarization (half-filling, $t'=0$). A, B
and C are defined in Fig. \ref{fig:TU_Phase-diagram}. Left: $\mathrm{Re}P^{\mathrm{ch}}(\mathbf{q},\omega=0)$.
Middle: $\mathrm{Re}P^{\mathrm{sp}}(\mathbf{q},\omega=0)$. Right:
$\mathrm{Im}\Sigma(\mathbf{k},i\omega_{0})$.}
\end{figure}

\begin{figure}
\begin{centering}
\includegraphics[scale=0.36]{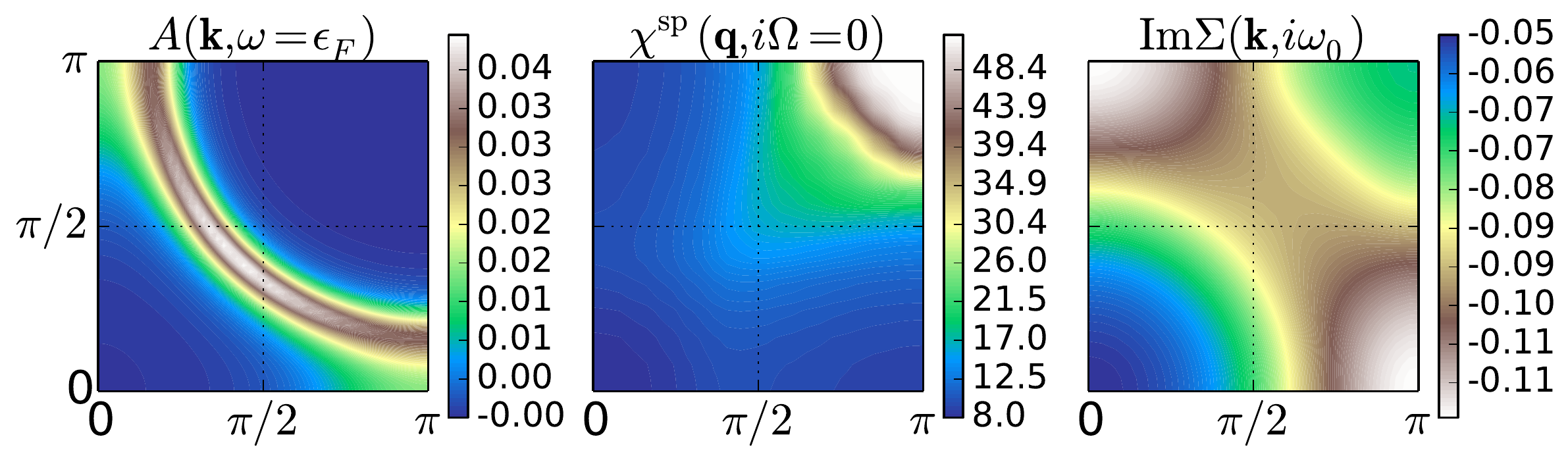}
\par\end{centering}

\protect\caption{(color online) From left to right: $A(\mathbf{k},\omega=0)$, $\chi(\mathbf{q},i\Omega=0)$
and $\mathrm{Im}\Sigma(\mathbf{k},i\omega_{0})$ in the doped case:
$U/D=1.8$, $t'=-0.4t$, $\beta D=96$, $\delta=10\%$.\label{fig:Doped-case}
($xyz$-decoupling)}

\end{figure}

Let us now apply the TRILEX method to the Hubbard model on a square
lattice. All energies are given in units of the half-bandwidth $D=4|t|$.
The Brillouin zone is discretized on a $64\times64$ momentum mesh.
We restrict ourselves to the paramagnetic normal phase.

In Fig. \ref{fig:TU_Phase-diagram}, we present the phase diagram
in the $(T,U)$ plane at half-filling. We obtain converged solutions
of the TRILEX scheme above a temperature denoted $T_{\mathrm{AF}}^{xyz}$
(resp. $T_{\mathrm{AF}}^{z}$) for the $xyz$-decoupling (resp. $z$-decoupling).
The evolution of the local vertex and self-energy (resp. lattice self-energy
and polarizations) is presented in Fig.\ref{fig:local_vertex} (resp.
Fig. \ref{fig:Momentum-dependence-Self-energies}) for the points
A, B and C of Fig. \ref{fig:TU_Phase-diagram}, in the $xyz$-decoupling.
At weak coupling (point A), the local vertex $\Lambda$ reduces to
the bare vertex $\lambda=1$ at large frequencies, up to numerical
noise (Fig. \ref{fig:local_vertex}a, upper panels). The spin polarization
(hence the spin susceptibility, see Eq.(\ref{eq:W_VS_chi})) becomes
sharply peaked at the AF wavevector $\mathbf{Q}=(\pi,\pi)$ (Fig.
\ref{fig:Momentum-dependence-Self-energies}, upper panels), reflecting
the nesting features of the Fermi surface. As a result, the self-energy
acquires a strong $\mathbf{k}$-dependence at $(\pi,0)$(Fig. \ref{fig:Momentum-dependence-Self-energies}),
but its local part is the same as the DMFT self-energy (Fig.\ref{fig:local_vertex}b).
At strong coupling (point C), the vertex becomes similar to the atomic
vertex (Fig. \ref{fig:local_vertex}a, lower panels). Furthermore,
the self-energy and polarization are weakly momentum-dependent (Fig.
\ref{fig:Momentum-dependence-Self-energies}, lower panels), in agreement
with cluster DMFT calculations; the self-energy of TRILEX is very
close to the DMFT self-energy (Fig. \ref{fig:local_vertex}b). Finally,
at intermediate coupling (point B), $\Lambda_{\mathrm{imp}}$ acquires
frequency structures which interpolate between A and C (Fig. \ref{fig:local_vertex}a,
middle panels), while $\Sigma$ is strongly momentum-dependent and
its local part departs from the DMFT self-energy (Fig. \ref{fig:local_vertex}b,
middle panels).

The temperature $T_{\mathrm{AF}}^{xyz/z}$ is determined by extrapolating
the inverse static AF susceptibility (Fig.\ref{fig:Inverse-susceptibility}
in Suppl. Mat.). It is reduced with respect to the Néel temperature
computed in DMFT \cite{Kunes2011} as a result of nonlocal fluctuations
beyond DMFT. Furthermore, $T_{\mathrm{AF}}^{xyz}\ll T_{\mathrm{AF}}^{z}$.
As a consequence of the apparent divergence in the spin susceptibility
at low temperatures (Fig. \ref{fig:Inverse-susceptibility}), caused
by a vanishing denominator of $W^{\eta}$ (Eq. \ref{eq:Dyson_W}),
we cannot obtain converged results in the close vicinity of and below
$T_{\mathrm{AF}}$. Whether we have an actual AF transition or finite
but very large correlation lengths (as seen \emph{e.g} in \cite{Schafer2014}),
could be decided by generalizing the present formalism to the symmetry-broken
phase. Contrary to cluster DMFT, the susceptibilities are not by-products
of the calculation, but directly enter the self-consistency loop through
$W^{\eta}$ (see Eq.(\ref{eq:W_VS_chi})). We thus cannot converge
paramagnetic solutions below an AF phase transition.

Let us now turn to the effect of doping. In Fig. \ref{fig:Doped-case},
we present results for $t'=-0.4t$, $\beta D=96$ and an intermediate
interaction strength ($U=1.8$, close to point B). The spectral function
displays Fermi arcs (Fig. \ref{fig:Doped-case}, left panel), as observed
in experiments \cite{Damascelli2003} and in cluster DMFT \cite{Jarrell2001,Kyung2004,Civelli2005,Kyung2006,Kyung2006a,Ferrero2008,Ferrero2009}.
Let us emphasize that this is obtained by solving a \emph{single-site}
quantum impurity problem, a far easier task than solving cluster impurities.
The Fermi arc is a consequence of the large static spin susceptibility
at the AF wavevector (Fig. \ref{fig:Doped-case}, middle panel), which
translates into a large imaginary part of the self-energy (Fig. \ref{fig:Doped-case},
right panel). The corresponding variation of the spectral weight on
the Fermi surface is rather mild due to the moderate correlation length
($\xi_{\mathrm{AF}}\sim2$ unit spacings) for these parameters.

Alternative self-consistency conditions are possible, \emph{e.g} $\chi_{\mathrm{loc}}^{\eta}=\chi_{\mathrm{imp}}^{\eta}$
instead of $W_{\mathrm{loc}}^{\eta}=W_{\mathrm{imp}}^{\eta}$ would
enforce sum rules on two-particle quantities that are key to preserving
the Mermin-Wagner theorem in \cite{Vilk1994,Katanin2009}. However,
this leads to a positive $\mathcal{U}^{\mathrm{sp}}(\tau)$ and hence
to a severe sign problem in the quantum Monte-Carlo at low temperatures.

In conclusion, we have presented the TRILEX formalism, which encompasses
long-range spin fluctuation effects and Mott physics in a unified
way. Like DMFT, it can be systematically controlled by extending it
to cluster schemes that interpolate between the single-site approximation
studied in this paper, and the exact solution of the model. We expect
that the convergence of the method as a function of the cluster size
will strongly depend on the decoupling channel and, when done in the
physically relevant channel, will be faster than cluster DMFT methods.
Furthermore, because the competition between spin fluctuations and
Mott physics can be described already at the single-site level, the
method may be a good starting point for correlated multiorbital systems
where spin fluctuations play an important role, like pnictides superconductors. 
\begin{acknowledgments}
We acknowledge useful discussions S. Andergassen, S. Biermann, M.
Ferrero, A. Georges, D. Manske, G. Misguich, J. Otsuki, A. Toschi.
We thank H. Hafermann for help with implementing the measurement of
the three-point correlation function. This work is supported by the
FP7/ERC, under Grant Agreement No. 278472- MottMetals.
\end{acknowledgments}

\bibliographystyle{apsrev4-1}
\bibliography{refs_trilex}

\clearpage{}

\appendix
\begin{widetext}

\part*{Supplementary Materials}

\setcounter{figure}{0} \renewcommand{\thefigure}{A.\arabic{figure}}

\section{Atomic Vertex}

In the atomic limit (single atomic site), one can compute the three-point
vertex exactly by writing its Lehmann representation. One gets the
following expression \cite{Ayral}: 

\begin{eqnarray}
\Lambda_{\mathrm{at}}^{\eta=\mathrm{ch/sp}}(i\omega_{n},i\Omega_{m}) & = & \frac{1}{1-U^{\eta}\chi_{\mathrm{at}}^{\eta}\delta_{m}}\Bigg(\frac{U^{2}/4}{i\omega_{n}\left(i\omega_{n}+i\Omega_{m}\right)}+1\nonumber \\
 &  & +\frac{U\beta\langle n_{\sigma}\rangle}{2}\left\{ 1-\frac{U^{2}}{4\left(i\omega_{n}\right)^{2}}\right\} \left\{ \tanh\left(\frac{\beta U}{4}\right)\mp1\right\} \delta_{m}\Bigg)\label{eq:Lambda_at}
\end{eqnarray}

where $\chi_{\mathrm{\mathrm{at}}}^{\eta=\mathrm{ch/sp}}\equiv\frac{\beta}{2}\frac{e^{\mp\beta U/4}}{\cosh(\beta U/4)}$
and at half-filling, $\langle n_{\sigma}\rangle=1/2$.

\section{Computation of the three-leg vertex\label{sec:Computation-of-the-three-leg-vertex}}

$\Lambda_{imp}$ is computed from the fermionic three-point correlation
function through the relation:
\begin{equation}
\Lambda_{\mathrm{imp}}^{\eta}(i\omega,i\Omega)=\frac{\tilde{\chi}_{\mathrm{imp,c}}^{\eta}(i\omega,i\Omega)}{G_{\mathrm{imp}}(i\omega)G_{\mathrm{imp}}(i\omega+i\Omega)\left(1-\mathcal{U}^{\eta}(i\Omega)\chi_{\mathrm{imp,c}}^{\eta}(i\Omega)\right)}\label{eq:Lambda_imp_from_chi}
\end{equation}
Here, the suffix ``c'' stands for ``connected'', namely:
\begin{equation}
\tilde{\chi}_{\mathrm{imp,c}}^{\eta}(i\omega,i\Omega)\equiv\tilde{\chi}_{\mathrm{imp}}^{\eta}(i\omega,i\Omega)-\beta G_{\mathrm{imp}}(i\omega)\langle n_{\mathrm{imp}}^{\eta}\rangle\delta_{i\Omega}\label{eq:chi_conn}
\end{equation}
while $\tilde{\chi}_{\mathrm{imp}}^{\eta}(i\omega,i\Omega)$ is defined
as $\tilde{\chi}_{\mathrm{imp}}^{\mathrm{ch}/\mathrm{sp}}(i\omega,i\Omega)=\tilde{\chi}_{\mathrm{imp}}^{\uparrow\uparrow}(i\omega,i\Omega)\pm\tilde{\chi}_{\mathrm{imp}}^{\uparrow\downarrow}(i\omega,i\Omega)$,
with:
\begin{equation}
\tilde{\chi}_{\mathrm{imp}}^{\sigma\sigma'}(i\omega,i\Omega)\equiv\iint_{0}^{\beta}d\tau d\tau'e^{i\omega\tau+i\Omega\tau'}\langle Tc_{\sigma}(\tau)c_{\sigma}^{\dagger}(0)n_{\sigma'}(\tau')\rangle\label{eq:def_chi3_tilde}
\end{equation}
$\chi_{\mathrm{imp}}^{\eta}(i\Omega)$ is the Fourier transform of
$\chi_{\mathrm{imp}}^{\eta}(\tau)\equiv\langle Tn^{\eta}(\tau)n^{\eta}(0)\rangle_{\mathrm{c}}$.

\section{Inverse spin susceptibility: Temperature evolution }

\begin{figure}[b]
\begin{centering}
\includegraphics[scale=0.32]{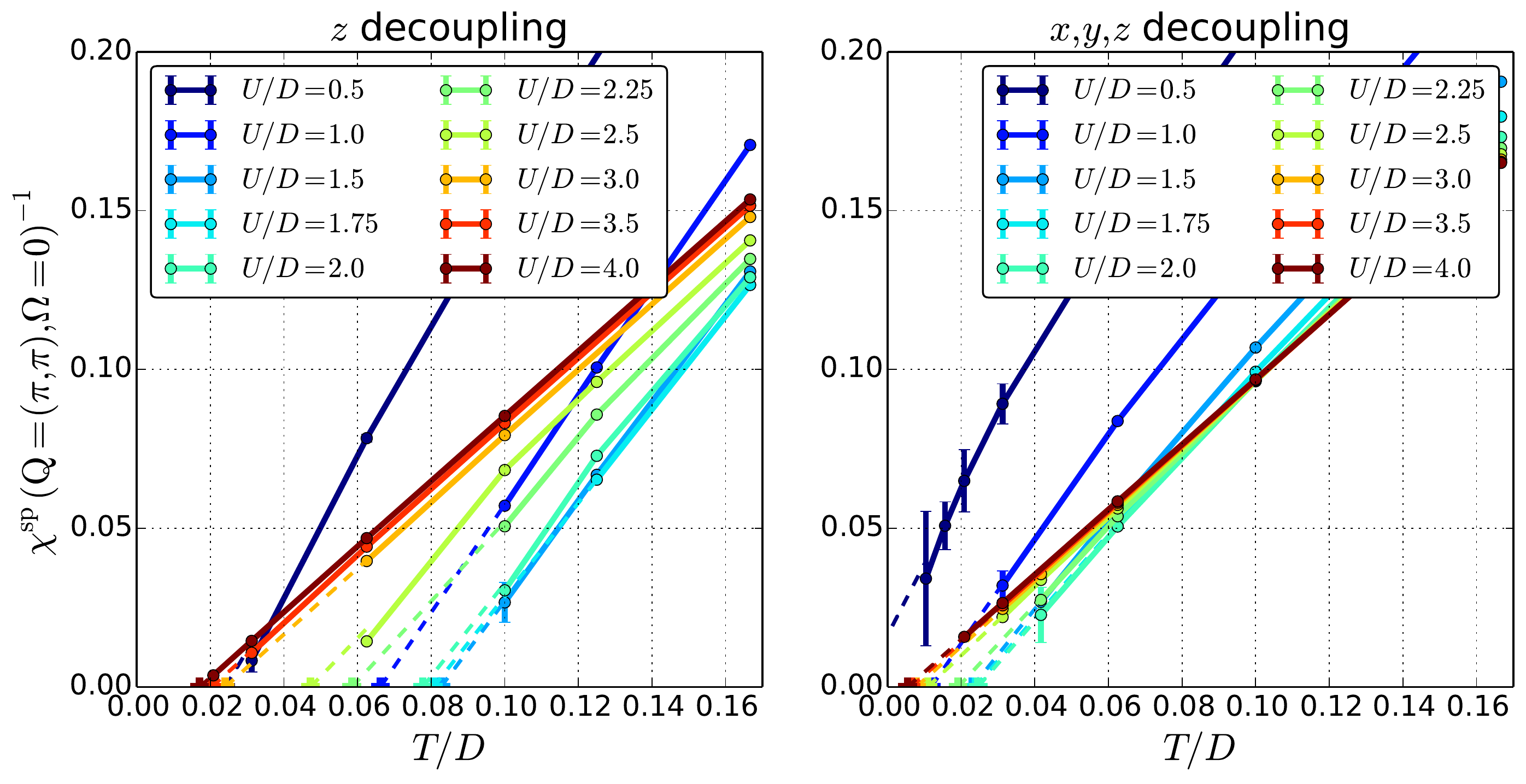}
\par\end{centering}

\protect\caption{(color online) Evolution of the inverse susceptibility with temperature
(half-filling, $t'=0$). The circles denote TRILEX computations, while
the crosses on the $x$-axis are the extrapolated points. The dashed
portion is the extrapolated line. Left: $z$-decoupling. Right: $xyz$-decoupling.
\label{fig:Inverse-susceptibility}}
\end{figure}

The inverse static AF spin susceptibility ($\chi^{\mathrm{sp}}(\mathbf{Q},i\Omega=0)^{-1}$,
obtained from $W^{\mathrm{sp}}(\mathbf{q},i\Omega)$ by Eq. (\ref{eq:W_VS_chi}))
decreases linearly with temperature, as shown in Fig. \ref{fig:Inverse-susceptibility}.
This allows to determine $T_{\mathrm{AF}}$ by extrapolation. The
resulting phase diagram is shown in Fig. \ref{fig:TU_Phase-diagram}.

\end{widetext}
\end{document}